\documentclass[prl,twocolumn,showpacs,preprintnumbers,amsmath,amssymb]{revtex4}

\usepackage{graphicx}
\usepackage{dcolumn}
\usepackage{bm}

\newcommand{\non}{\nonumber}
\newcommand{\rr}{{\bm r}}
\newcommand{\ee}{{\bm e}}
\newcommand{\hS}{\hat{S}}
\newcommand{\bms}{\hat{\bm s}}
\newcommand{\muB}{\mu_{\text{B}}}
\newcommand{\cdd}{c_{\rm dd}}
\newcommand{\vdd}{\hat{v}_{\rm dd}}
\newcommand{\Beff}{B_{\rm eff}}
\newcommand{\Bext}{B_{\rm ext}}
\newcommand{\aho}{a_{\rm ho}}

\begin{document}

\preprint{APS/123-QED}


\title{Einstein--de Haas Effect in Dipolar Bose--Einstein Condensates}
\author{Yuki Kawaguchi$^1$}
\author{Hiroki Saito$^1$}
\author{Masahito Ueda$^{1,2}$}
\affiliation{$^1$Department of Physics, Tokyo Institute of Technology,
2-12-1 Ookayama, Meguro-ku, Tokyo 152-8551, Japan \\
$^2$ERATO, Japan Science and Technology Corporation (JST), Saitama
332-0012, Japan
}
\date{\today}

\begin{abstract}
The general properties of the order parameter of a dipolar spinor Bose--Einstein condensate are discussed based on the symmetries of interactions.
An initially spin-polarized dipolar condensate is shown to dynamically generate a non-singular vortex via spin-orbit interactions, a phenomenon reminiscent of the Einstein--de Haas effect in ferromagnets. 
\end{abstract}


\maketitle

The realization of a Bose--Einstein condensate (BEC) of $^{52}$Cr~\cite{Griesmaier2005,Stuhler2005} marks a major development in degenerate quantum gases in that the interparticle interaction via magnetic dipoles in this BEC is much larger than those in other spinor BECs of alkali atoms.
The long-range nature and anisotropy of the dipolar interaction pose challenging questions concerning the stability and superfluidity of the BEC~\cite{Yi2001,Goral2000,Santos2000,Goral2002,Baranov2002,Santos2003,ODell2004,Eberlein2005}.
The ground state of the $^{52}$Cr atom has a total electronic spin of three and a nuclear spin of zero, and therefore the $^{52}$Cr BEC has seven internal degrees of freedom. 
The interplay between dipolar and spinor interactions makes the order parameter of this system highly non-trivial~\cite{Pu2001,Gross2002,Yi2004}.
Moreover, the dipole interaction couples the spin and orbital angular momenta so that an initial magnetization of the system causes the gas to rotate mechanically (Einstein--de Haas effect~\cite{Einstein-deHaas}) or, conversely, solid-body rotation of the system leads to its magnetization (Barnett effect~\cite{Barnett}).

This Letter investigates the Einstein--de Haas and Barnett effects in a spin-3 BEC system.
We discuss the symmetry of the order parameter of a dipolar spinor BEC and
study the dynamic formation of spin textures using numerical simulations of the seven-component nonlocal mean-field theory,
which takes into account short-range (van der Waals) interactions and magnetic dipole--dipole interactions subject to a trapping potential and an external magnetic field.

We first consider general properties of the order parameter by discussing two fundamental symmetries of the dipolar interaction between the magnetic dipole moments ${\bm \mu}_1=g\muB\bms_1$ and ${\bm \mu}_2=g\muB\bms_2$, where $g$ is the electron $g$-factor, $\muB$ is the Bohr magneton, and $\bms_1$ and $\bms_2$ are the spin operators. The interaction between the magnetic dipoles located at $\rr_1$ and $\rr_2$ is described by
\begin{align}
\vdd(\rr_{12}) = \cdd\frac{(\bms_1\cdot\bms_2)-3(\bms_1\cdot\ee_{12})(\bms_2\cdot\ee_{12})}{r_{12}^3},
\label{eq:vdd}
\end{align}
where
$\rr_{12}\equiv\rr_1-\rr_2$, $\ee_{12}\equiv\rr_{12}/r_{12}$, and $\cdd=\mu_0(g\muB)^2/4\pi$
with $\mu_0$ being the magnetic permeability of the vacuum.
The dipole interaction is invariant under simultaneous rotation in spin and coordinate spaces about an arbitrary axis, say the $z$-axis, so that 
the projected total angular momentum operator $\hat{S}_z+\hat{L}_z$ on that axis  commutes with $\vdd$, where $\hat{\bm S}=\bms_1+\bms_2$ is the total spin operator and $\hat{\bm L}$ is the relative orbital angular momentum operator.
Another symmetry of the dipolar interaction is the invariance under the transformation ${\sf P}_z e^{-i\pi \hS_z}$, 
where ${\sf P}_z : (x,y,z)\to(x,y,-z)$ 
and $e^{-i\pi \hS_z} : (\hS_x, \hS_y, \hS_z)\to(-\hS_x,-\hS_y,\hS_z)$.
Thus, the eigenvalues of the following operators are conserved by the dipole interaction:
\begin{align}
\hat{S}_z+\hat{L}_z \hspace{5mm}{\rm and}\hspace{5mm} {\sf P}_z~e^{-i\pi \hat{S}_z}.
\label{eq:symmetry}
\end{align}
A crucial observation is that these operators also commute with the short-range interactions. Thus if the confining potential is axisymmetric, the simultaneous eigenfunctions of the two operators~\eqref{eq:symmetry} can serve to classify the two-body wave function.

Constructing a many-body wave function by directly applying these symmetry considerations is quite complicated 
since the system has many degrees of freedom.
However, substantial simplification can be achieved by considering the case of a ferromagnet in which dipole moments are localized at lattice sites
and thus the degrees of freedom of the system are reduced to the center-of-mass motion and the solid-body rotation around it.
Consequently, spin relaxation of the system leads to a solid-body rotation of the ferromagnet --- the Einstein--de Haas effect~\cite{Einstein-deHaas}.
An analogous consideration can be applied to a BEC because almost all atoms occupy a single-particle state and therefore
the degrees of freedom of the system can be represented by those of the order parameter.
We may thus expect the Einstein--de Haas effect to emerge in a dipolar spinor BEC system.

In general, the order parameter of a BEC can be defined as the eigenfunction corresponding to the macroscopic eigenvalue of the reduced single-particle density operator.
Let the order parameter be denoted as $\psi_\alpha(\rr)$ with a norm assumed to be $N$, the number of condensate atoms.
Here, $\alpha$ represents the magnetic sublevels of the atoms.
It follows from the above symmetry considerations of the dipolar interaction that
the order parameter of a dipolar spinor BEC in an axisymmetric system can be 
classified by the eigenvalues of the operators $\hat{s}^z + \hat{l}^z$ and ${\sf P}_z \exp(-i\pi \hat{s}^z)$
corresponding to the operators~\eqref{eq:symmetry},
where $\hat{s}^z$ is the $z$ component of the spin matrix and $\hat{l}^z\equiv-i(\partial/\partial\phi)$.
The simultaneous eigenstate of $\hat{s}^z + \hat{l}^z$ and ${\sf P}_z \exp(-i\pi \hat{s}^z)$
with eigenvalues $J$ and $p$, respectively, is given by
\begin{align}
\psi_\alpha(r,\phi,z,t)=e^{i(J-\alpha)\phi}\eta_{\alpha Jp}(r,z,t),
\label{eq:spinorOP}
\end{align}
where $J$ is an integer and
$\eta_{\alpha Jp}$ is a complex eigenfunction of ${\sf P}_z$ satisfying ${\sf P}_z \eta_{\alpha Jp} = p (-1)^\alpha \eta_{\alpha Jp}$ with $p=\pm 1$.
A key point of Eq.~\eqref{eq:spinorOP} is that it includes the spin-dependent phase factor,
since the dipolar interaction couples the spin with the orbital angular momentum.

We now consider the dynamic formation of spin textures in a dipolar BEC.
The order parameter obeys the following set of the nonlocal Gross--Pitaevskii equations:
\begin{align}
i\hbar\frac{d\psi_\alpha(\rr)}{dt} 
&= \left[-\frac{\hbar^2\nabla^2}{2M} + g\muB \Bext\alpha + U_{\rm trap}(\rr) \right]\psi_{\alpha}(\rr)\non\\
&+ \sum_{\beta\alpha'\beta'}\sum_{S=0}^{2s}
g_S\langle \alpha\beta|\mathcal{P}_S|\alpha'\beta'\rangle \psi^*_{\beta}(\rr)\psi_{\alpha'}(\rr)\psi_{\beta'}(\rr)\non\\
&+ \sum_{\mu=x,y,z}\sum_\beta \Beff^\mu(\rr) (g\muB s^\mu_{\alpha\beta})\psi_\beta(\rr),
\label{eq:GP}
\end{align}
where $M$ is the atomic mass,
$\Bext$ is the external magnetic field in the $z$ direction, 
and $U_{\rm trap}(\rr)$ is the trapping potential.
We assume an optical trap so that all spin components experience the same trapping potential.
The second line in Eq.~\eqref{eq:GP} represents the short-range interaction, where
the strength of the interaction is characterized by the {\it s}-wave scattering length $a_S$ for a total spin $S$ of a pair of atoms with spin $s$ as $g_S=4\pi\hbar^2 a_S/M$.
The operator $\mathcal{P}_S$ projects the wave function into the Hilbert
space with a total spin $S$ and is represented in terms of the Clebsch--Gordan coefficients 
$\langle s\alpha s\beta|SM_S \rangle$ as
$\langle \alpha\beta|\mathcal{P}_S|\alpha'\beta'\rangle = \sum_{M_S=-S}^S\langle s\alpha s\beta|SM_S\rangle \langle SM_S | s\alpha's\beta'\rangle$~\cite{Ho1998}.

The last line in Eq.~\eqref{eq:GP} represents the dipole--dipole interaction,
where $s^{x,y,z}$ are the spin matrices, and
\begin{align}
\Beff^\mu(\rr) \equiv 
 \frac{\cdd}{g\muB} \sum_\nu \int d\rr' &
 \frac{\delta_{\mu\nu}-3 e^\mu e^\nu}{|\rr-\rr'|^3}\non\\
 &\sum_{\alpha'\beta'}\psi_{\alpha'}^*(\rr') s^\nu_{\alpha'\beta'}\psi_{\beta'}(\rr')
\label{eq:Beff}
\end{align}
is the effective magnetic field at $\rr$ produced by the surrounding magnetic dipoles, with $\ee\equiv (\rr-\rr')/|\rr-\rr'|$.
Calculating the time derivative of $S^\mu(\rr)=\sum_{\alpha\beta}\psi^*_\alpha(\rr) s^\mu_{\alpha\beta}\psi_\beta(\rr)$,
we find that, apart from the spinor interactions, ${\bm S}(\rr)$ behaves like a classical spin
and undergoes Larmor precession around the effective local magnetic field ${\bm B}_{\rm eff}(\rr)+\Bext \hat{z}$.
Hence, spin flip occurs in the region where $\Beff^{x,y}(\rr)\neq 0$.
In a homogeneous infinite system, the effective field is completely canceled in a polarized BEC
and spin flip does not occur.
Therefore, the Einstein--de Haas effect in an initially fully spin-polarized BEC is unique to non-uniform systems.

We now examine the spin dynamics of the Einstein--de Haas effect in a spin-3 $^{52}$Cr BEC system.
We consider a stable spin-polarized BEC in the lowest magnetic sublevel $\alpha=-3$, produced
in a strong magnetic field, as in the experiments of Refs.~\cite{Griesmaier2005,Stuhler2005}.
We then suddenly decrease the magnetic field to $\Bext$.
The initial state can be calculated by the imaginary-time propagation method in the subspace of $\psi_{-3}(\rr)$.
We have performed three-dimensional simulations of Eq.~\eqref{eq:GP} with seven spin components by using the Crank--Nicolson method.
The scattering lengths of $^{52}$Cr are reported to be $a_6=112, a_4=58$, and $a_2=-7$ in units of the Bohr radius~\cite{Werner2005}.
The value of $a_0$ is unknown and we estimate it using
the van der Waals coefficient $C_6$ given in Ref.~\cite{Werner2005}, obtaining $a_0\sim(C_6 M/m_{\text{e}})^{1/4}=91$, where $m_{\text{e}}$ is the electron mass.
Since we are interested in the spin dynamics of a fully spin-polarized state,
the short-range interaction is dominated by $a_6$ and $a_4$~\cite{note}.

We assume $N=10^5$ atoms trapped in an axisymmetric potential $U_{\rm trap}(\rr)=(1/2)M\omega^2(r^2+z^2/\lambda^4)$ with $\omega=2\pi\times820$~Hz.
The typical ratio of the dipolar interaction to the short-range interaction is $s^2\cdd/g_6\simeq0.03$ with $s=3$.
In a spherical trap, the number density at the trap center is $n\simeq7\times10^{20}$~m$^{-3}$
and the dipole--dipole interaction energy $s^2\cdd n$ becomes of the same order of magnitude as the Zeeman energy for $\Bext\simeq 0.1$~mG.
Hence, the spin-flip rate becomes significant for $|\Bext|\lesssim1$~mG.
In the following,
we first consider $\Bext=0$ and $\lambda=1$, and 
then discuss the effects of the external magnetic field and the trap geometry on the spin dynamics.

\begin{figure}
\includegraphics[width=0.95\linewidth]{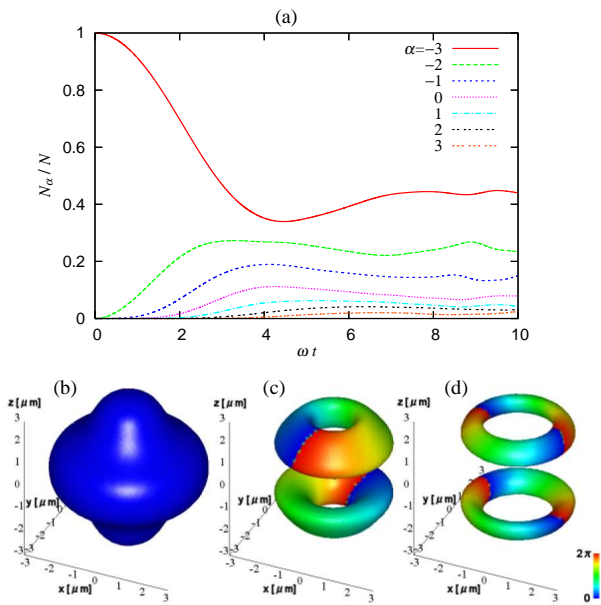}
\caption{(Color) (a) Relative population $N_\alpha/N=\int d\rr |\psi_\alpha|^2/N$ of each magnetic sublevel $\alpha$ versus $\omega t$ for $\Bext=0$ and $\lambda=1$.
(b)--(d) Isopycnic surfaces of (b) $\psi_{-3}$, (c) $\psi_{-2}$, and (d) $\psi_{-1}$ at $\omega t=2$, where $|\psi_\alpha|^2\aho^3/N= 0.0001$ for (b) and (c) and $5\times10^{-5}$ for (d).
The color on the surfaces represents the phase of the order parameter (refer to scale at right).
}
\label{fig:1}
\end{figure}

Figure~\ref{fig:1} shows the results for $\Bext=0$ and $\lambda=1$.
In Fig.~\ref{fig:1}(a), we plot the population of each spin state $N_\alpha/N\equiv\int d\rr |\psi_\alpha|^2/N$ as a function of $\omega t$. 
The figure shows that $N_{-2}/N$ first increases rapidly and then the components with $\alpha\geq -1$ begin to increase.
Figures~\ref{fig:1}(b), (c), and (d) show three-dimensional plots of $\psi_{-3}$, $\psi_{-2}$, and $\psi_{-1}$, respectively, at $\omega t=2$,
where $|\psi_\alpha|^2$ is scaled by $N/\aho^3$ with $\aho$ being the harmonic oscillator length $\sqrt{\hbar/2M\omega}$.
The order parameters show the symmetries of Eq.~\eqref{eq:spinorOP}:
$\psi_{-2}$ has a phase factor $e^{-i\phi}$ and a node plane at $z=0$, and
$\psi_{-1}$ has a phase factor $e^{-2i\phi}$ and has reflection symmetry with respect to the $z=0$ plane.
The other spin components have double-ring shapes similar to that of Fig.~\ref{fig:1}(d)
and their phase relationships satisfy Eq.~\eqref{eq:spinorOP} with $J=-3$ and $p=-1$.
Beyond $\omega t=5$, the spinor order parameter of the system develops a complicated structure.

\begin{figure}
\includegraphics[width=0.9\linewidth]{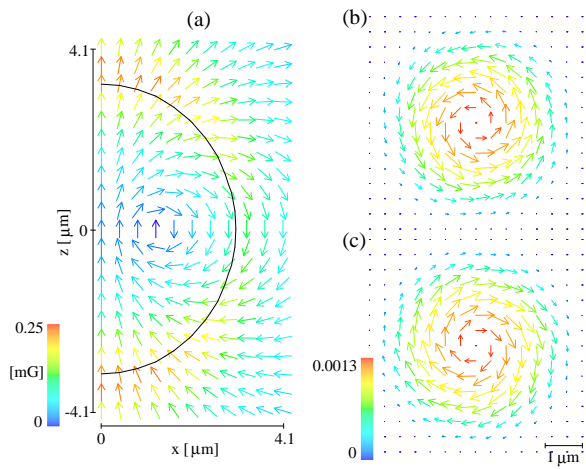}
\caption{(Color) (a) Dipole field at $t=0$ for $\lambda=1$.
The solid line shows the isopycnic curve at $|\psi_{-3}|^2\aho^3/N = 0.0001$.
The color of the arrows denotes the magnitude of the field.
(b),(c) Spin configurations on the (b) $z=2~\mu$m plane and (c) $z=-2~\mu$m plane at $\omega t=2$ for $\Bext=0$ and $\lambda=0$.
The length of the arrows represents the magnitude of the spin vector projected on the $xy$-plane and the color shows $|{\bm S}|\aho^3/N$.
Note that the spins tilt in a direction perpendicular to ${\bm B}_{\rm eff}(\rr)$.
}
\label{fig:2}
\end{figure}

The effective magnetic field, Eq.~\eqref{eq:Beff}, at $\omega t=0$ and 
the spin vector ${\bm S}(\rr)$ at $\omega t=2$ are plotted in Fig.~\ref{fig:2}.
The whirling patterns of the spin texture in Figs.~\ref{fig:2}(b) and (c) are due to Larmor precession
around the local magnetic field, shown in Fig.~\ref{fig:2}(a).
Since the local magnetic field points outward for $z>0$ and inward for $z<0$,
the directions of the whirlpools in Figs.~\ref{fig:2}(b) and (c) are opposite.
Topological spin textures in spinor BECs have been observed in a spin-1/2 system~\cite{Matthews1999} and in a spin-1 system~\cite{Leanhardt2003}.
We note that the spin textures in Figs.~\ref{fig:2}(b) and (c) are generated spontaneously due to the intrinsic dipole interaction,
in contrast to those in Refs.~\cite{Matthews1999, Leanhardt2003}, which are generated by external forces.

\begin{figure}
\includegraphics[width=\linewidth]{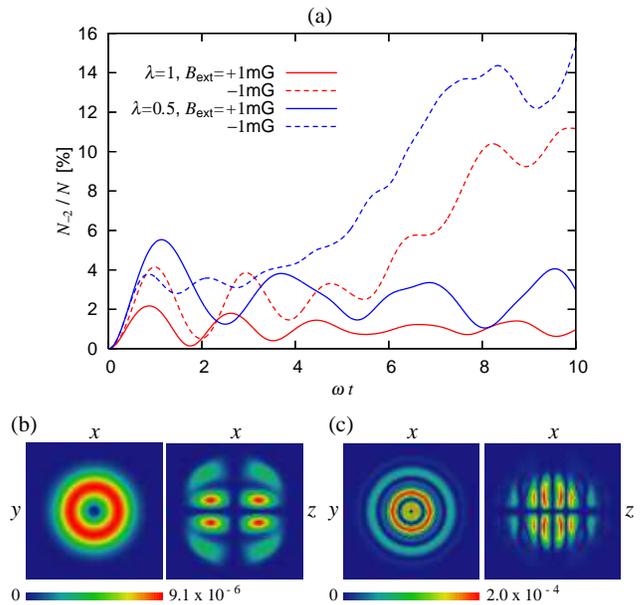}
\caption{(Color) (a) Population of $\alpha=-2$ sublevel versus $\omega t$ for $\Bext=\pm1$~mG and $\lambda=1$ and 0.5.
(b),(c) Cross sections of $|\psi_{-2}|^2\aho^3/N$ with $\lambda=1$ at $\omega t=8$ for (b) $\Bext=1$~mG and (c) $\Bext=-1$~mG.
The cross sections are at $z=0.7~\mu$m (left) and $y=0~\mu$m (right).
The size of each panel is $8.2\times8.2~\mu$m.
}
\label{fig:Bdep}
\end{figure}

When the external magnetic field $\Bext$ is applied in the positive $z$ direction and is much stronger than the dipole field, the spin angular momentum should be conserved because of energy conservation and spin flipping is suppressed.
When $\Bext<0$, spin flip can occur by converting Zeeman energy to kinetic energy.
These behaviors are demonstrated in Fig.~\ref{fig:Bdep}(a), which shows the time evolution of $N_{-2}/N$ for $\Bext=\pm1$~mG.
Figures~\ref{fig:Bdep}(b) and (c) show cross sections of $|\psi_{-2}|^2 \aho^3/N$ for $\Bext=\pm 1$~mG at $\omega t=8$ and $\lambda=1$.
When $\Bext>0$, $\psi_{-2}$ oscillates in time between the structures of Fig.~\ref{fig:1}(c) and Fig.~\ref{fig:Bdep}(b).
These structures derive from the symmetry of the dipole interaction, which can be expressed in terms of rank-2 spherical harmonics $Y_{2m}$.
The dipole field produced by an approximately spherical distribution of
$\psi_{-3} \sim Y_{00}(\rr)$ induces a $Y_{2-1}(\rr)$ term in $\psi_{-2}$ (Fig.~\ref{fig:1}(c)),
which in turn affects itself and induces a linear combination of $Y_{2-1}(\rr)$ and $Y_{4-1}(\rr)$, resulting in Fig.~\ref{fig:Bdep}(b).
Therefore, the structure in Fig.~\ref{fig:Bdep}(b) manifests as a secondary effect of the dipole--dipole interaction.
In the case of $\Bext=-1$~mG, a similar structure as in Fig.~\ref{fig:Bdep}(b) appears for $\omega t \lesssim 4$.
However, as time advances, the domain structure develops as shown in Fig.~\ref{fig:Bdep}(c).
The domain size becomes smaller as $|\Bext|$ increases.
\begin{figure}
\includegraphics[width=0.7\linewidth]{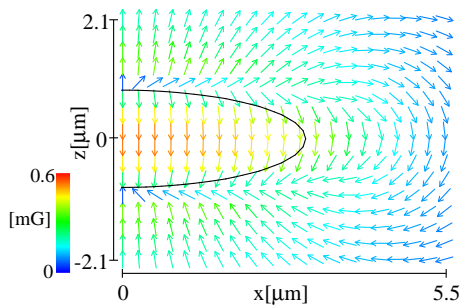}
\caption{(Color) Dipole field at $t=0$ for $\lambda=0.5$.
The solid line shows the isopycnic curve at $|\psi_{-3}|^2\aho^3/N = 0.0005$.
Note that the sign of $\Beff^z(\rr)$ in the condensate is opposite to that in Fig.~2(a).
}
\label{fig:pancake}
\end{figure}

Finally, we discuss the geometry dependence of the dynamics.
Figure~\ref{fig:pancake} shows the dipole field ${\bm B}_{\rm eff}(\rr)$ at $t=0$ for $\lambda=0.5$.
Compared with Fig.~\ref{fig:2}(a), the $z$ component of the effective magnetic field is inverted around the center of the condensate.
The transfer of atoms from the spin component $\alpha=-3$ to $\alpha\geq -2$ is due to Larmor precession caused by $\Beff^{xy}$,
and it occurs most efficiently where the local magnetic field in the $z$ direction $\Beff^z+\Bext$ vanishes.
The sign of such an optimized $\Bext$ for spherical traps is opposite to that for pancake-shaped traps,
as can be inferred from Fig.~\ref{fig:2}(a) and Fig.~\ref{fig:pancake}.
This fact is reflected in Fig.~\ref{fig:Bdep}(a) 
as the difference in the field dependence of the initial peaks.
Since the $z$ component of the dipole field is positive for most of the condensate when $\lambda=1$ (Fig.~\ref{fig:2}(a)),
the initial peak in Fig.~\ref{fig:Bdep}(a) is larger for $\Bext=-1$~mG than for $\Bext=1$~mG.
The relation between the initial peak and $\Bext$ is opposite for $\lambda=0.5$, since the $z$ component of the dipole field is mostly negative in the condensate (Fig.~\ref{fig:pancake}).
In the case of a cigar-shaped BEC, the qualitative behavior is the same as for the spherical trap.

In conclusion, 
we have shown that the Einstein--de Haas effect occurs in dipolar spinor Bose--Einstein condensates and that a non-singular vortex appears from an initially spin-polarized condensate.
In a low magnetic field ($\sim 1$~mG)
the fraction of spin-flipped atoms ($\sim5$~\%) is large enough to be observed in Stern--Gerlach experiments.
The spin-relaxation processes produce various vortex structures, depending on the external magnetic field.

\begin{acknowledgments}
This work was supported by Grant-in-Aids for Scientific Research (Grant
No.\ 17740263, No.\ 17071005, and No.\ 15340129) and by a 21st Century COE
program at Tokyo Tech ``Nanometer-Scale Quantum Physics'' from the
Ministry of Education, Culture, Sports, Science and Technology of Japan.
YK acknowledges support by a  Fellowship Program of the Japan Society for 
Promotion of Science (Project No.\ 160648).
MU acknowledges support by a CREST program of the JST.
\end{acknowledgments}

{\it Note added.}
--- Very recently, a preprint~\cite{condmat0510634} has appeared which also discusses the Einstein--de Haas effect in a dipolar BEC.


\end{document}